\documentclass[a4paper,11pt]{article}
\usepackage{pos}
\usepackage{tabularx}
\usepackage{amsmath}
\usepackage{tikz}

\usepackage{titlesec}
\titlespacing{\paragraph}{%
  0pt}{
  0.5\baselineskip}{
  1em}

\title{Tensor-network toolbox for probing dynamics of non-Abelian gauge theories}

\author*[a,b]{Emil Mathew}
\author[c,d]{Navya Gupta}
\author[e]{Saurabh V. Kadam}
\author[d,f]{Aniruddha Bapat}
\author[f]{Jesse Stryker}
\author[c,d]{Zohreh Davoudi}
\author[a,b]{Indrakshi Raychowdhury}

\affiliation[a]{Department of Physics,
BITS Pilani K K Birla Goa Campus,
Zuarinagar, Goa 403726, India}
\affiliation[b]{Center for Research in Quantum Information and Technology (CRQIT), BITS Pilani K K Birla Goa Campus, Zuarinagar, Goa 403726, India}
\affiliation[c]{Maryland Center for Fundamental Physics and Department of Physics, University of Maryland, College Park, MD 20742, USA.}
\affiliation[d]{Joint Center for Quantum Information and Computer Science, University of Maryland, College Park, Maryland 20742, USA.}
\affiliation[e]{InQubator for Quantum Simulation (IQuS), Department of Physics, University of Washington, Seattle, WA 98195, USA}
\affiliation[f]{Physics Division, Lawrence Berkeley National Laboratory, Berkeley, CA 94720, USA.}

\emailAdd{p20210036@goa.bits-pilani.ac.in}

\abstract{Tensor-network methods enable probing dynamics of strongly interacting quantum many-body systems, including gauge theories, via Hamiltonian simulation, hence bypassing sign problems. They also have the potential to inform efficient quantum-simulation algorithms of the same theories. We develop and benchmark a matrix-product-state ansatz for the SU(2) lattice gauge theory using the loop-string-hadron formulation. This formulation has been demonstrated to be advantageous in Hamiltonian simulation of non-Abelian gauge theories. It is applicable to both SU(2) and
SU(3) gauge groups, to periodic and open boundary conditions, and to 1+1 and higher dimensions. In this work, we report on progress in computing static and dynamical observables in a SU(2) gauge theory in (1+1)D, pushing the boundary of existing studies. 
}

\FullConference{The 41st International Symposium on Lattice Field Theory (LATTICE2024)\\
 28 July - 3 August 2024\\
Liverpool, UK\\}

\begin{document}
\maketitle

\section{Introduction}
Tensor networks (TNs) are promising sign-problem-free tools that enable studies of static and dynamic properties of quantum many-body systems~\cite{white1992density,fannes1992finitely,ostlund1995thermodynamic,verstraete2004density,schollwock2011density,cirac2020matrix,meurice2020tensor,banuls2020review}. In the context of high-energy and nuclear physics, much effort has been directed at lattice gauge theories (LGTs), both Abelian~\cite{byrnes2002density,tagliacozzo2011entanglement,rico2014tensor,buyens2014matrix,silvi2014lattice,tagliacozzo2014tensor,zohar2015formulation,rigobello2021entanglement,frias2022light,canals2024tensor,banuls2023quantum,funcke2023exploring,magnifico2024tensor,banuls2024parton,belyansky2024high} and non-Abelian gauge theories \cite{kuhn2015non,banuls2017efficient,sala2018variational,silvi2019tensor,rigobello2023hadrons,hayata2024dense}, in 1+1 and higher dimensions ~\cite{Bender:2020jgr,emonts2020variational,magnifico2021lattice,robaina2021simulating,Emonts:2022yom,Cataldi:2023xki,Kelman:2024exo}, see also Refs.~\cite{banuls2018tensor,banuls2020review,meurice2022tensor,Magnifico:2024eiy} for reviews.

This work reports a tensor-network study of the SU(2) LGT in (1+1)D, employing a recently-developed theoretical formulation~\cite{Raychowdhury:2019iki}. This LGT has been extensively studied in Refs.~\cite{kuhn2015non,banuls2017efficient,sala2018variational}, where different formulations of the SU(2) gauge theory~\cite{davoudi2021search}, namely truncated angular-momentum basis and purely fermionic basis, were used to compute static and dynamic quantities. The static quantities of interest comprise ground-state energy estimation via the density-matrix renormalization group (DMRG) and its continuum-limit extrapolation for various model parameters. In contrast, the dynamical computation explores string-breaking dynamics in the presence of static and dynamical charges on the lattice. (See Refs.~\cite{Cochran:2024rwe,De:2024smi,Gonzalez-Cuadra:2024xul,Ciavarella:2024lsp,Crippa:2024hso,Liu:2024lut} for recent quantum-simulation experiments of string breaking in Abelian models). While Refs.~\cite{kuhn2015non,sala2018variational} observe string-breaking dynamics in the SU(2) LGT, they are either constrained to small Hilbert-space truncation cutoffs and lattice sizes (Ref.~\cite{kuhn2015non}) or to the evolution of only static charges (Ref.~\cite{sala2018variational}). 

This work is aimed at extending existing results, setting the stage for studies of other non-equilibrium processes, such as post-collision phenomena, in the SU(2) gauge theory. Specifically, we simulate the dynamics of string breaking by exploring sufficiently long strings composed of dynamical fermions embedded in sufficiently large lattice volumes to enable approaching the continuum limit. We also control the gauge-boson-truncation effects, reaching larger cutoffs than previously accessible. As a result, we observe richer phenomenology compared to previous studies. Our work is enabled by a gauge-invariant reformulation of the Kogut-Susskind LGT~\cite{kogut1975hamiltonian} based on loop, string, and hadron degrees of freedom~ \cite{Raychowdhury:2019iki}, resulting in a simpler Abelianized theory. 

This proceedings is organized as follows: Section~\ref{sec:LSH} briefly reviews the loop-string-hadron formalism, followed by Sec.~\ref{sec:LSH_network}, where the tensor-network framework for the LSH formulation is introduced. In Sec.~\ref{sec:static_properties}, we compute the ground-state energy along with the effect of external static charges, i.e., the static potential. 
Section~\ref{sec:dynamic_properties} contains a study of the quenched dynamics of dynamical strings and the string-breaking phenomenon. We conclude in Sec.~\ref{sec:conclusion}.

\section{Loop-string-hadron formulation}
\label{sec:LSH}
The loop-string-hadron (LSH) formulation replaces the gauge-invariant degrees of freedom, such as Wilson loops/lines, by local snapshots of the same. These snapshots are defined using a gauge-invariant and orthonormal LSH basis, which in (1+1)D is characterized by a set of three integers $\{n_l, n_i, n_o\}$.
$n_l \in \mathbb{Z}^+\cup \{0\}$ is a bosonic quantum number while $n_i,n_o \in \{0,1\}$ are fermionic quantum numbers. The global state is a tensor-product state associated with these quantum numbers at each site $r$ of the spatial lattice, i.e., $\bigotimes_{r=1}^{N} |n_l,n_i,n_o\rangle_r$, glued together via an Abelian Gauss law (AGL), $N_L(r)=N_R(r+1)$, where $N_L (r) \coloneq \big[ n_l+n_o(1-n_i)\big]\Big|_r$ and $N_R (r) \coloneq \big[ n_l+n_i(1-n_o)\big]\Big|_r$, $\forall r$.
For a more detailed exposition of the LSH formalism, see Ref.~\cite{Raychowdhury:2019iki}.

The Hamiltonian in the LSH formulation can be written as:
\begin{align}
\label{eq:H_LSH}
\hat H^{(\rm LSH)}&=\frac{g^2a}{4}\sum_{r=1}^{N-1}\Bigg[ \frac{\hat{N}_L(r)}{2}
\left( \frac{\hat{N}_L(r)}{2}+1 \right) 
+ \Big(\hat{N}_L(r) \leftrightarrow \hat{N}_R(r+1) \Big)
\Bigg] + m\sum_{r=1}^{N} (-1)^r(\hat n_i(r)+\hat n_o(r))\nonumber\\
&+ \frac{1}{2a}\sum_{r=1}^{N-1} \Bigg\{\frac{1}{\sqrt{\hat{N}_L(r)+1}}\Big[\hat{\mathcal{S}}_\text{out}^{++}(r)\hat{\mathcal{S}}_\text{in}^{+-}(r+1)
+ \hat{\mathcal{S}}_\text{out}^{+-}(r)\hat{\mathcal{S}}_\text{in}^{--}(r+1)\Big]\frac{1}{\sqrt{\hat{N}_R(r+1)+1}}+ \text{H.c}.\Bigg\}.
\end{align}
Here, $g$ is the coupling strength, $m$ is the mass of the staggered fermions, and $a$ is the lattice spacing. The operators used in the Hamiltonian are defined in Table~\ref{tab:lsh_dict}. In this work, we apply open boundary conditions (OBCs) but our formalism is also applicable to periodic boundary conditions (PBCs).
\setlength{\textfloatsep}{0pt plus 16.0pt minus 2.0pt}
\begin{table}[t!]
        \includegraphics[scale=0.89]{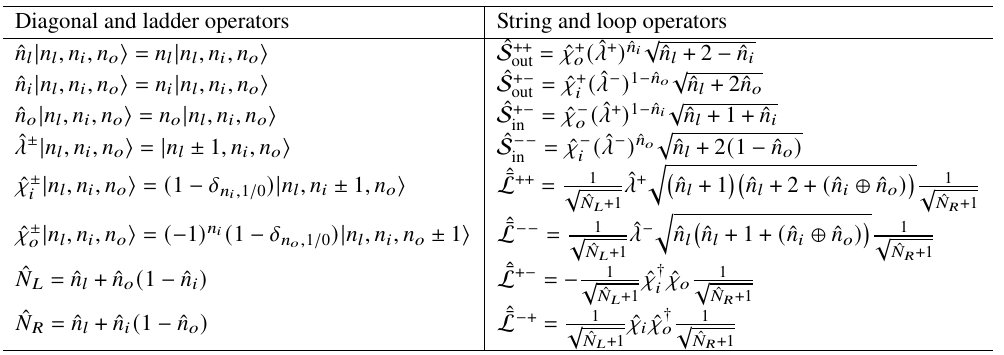}
        \caption{The list of relevant operators for defining the LSH Hamiltonian and dynamical strings in (1+1)D. Note that $\hat{\lambda}^{-}|0, n_i, n_o\rangle = 0$ and the symbol $\oplus$ is used to denote addition modulo 2.
        These definitions are valid for a single site; when embedded in a lattice, the operators carry a position label $r$ and the $\chi(r)$ definitions must be endowed with additional structure to ensure proper anticommutation relations \cite{Raychowdhury:2019iki}.
        }\label{tab:lsh_dict}
\end{table}

\section{Tensor-network framework for the LSH formulation}
\label{sec:LSH_network}
A matrix-product-state (MPS) ansatz for a (1+1)D system with $N$ spatial sites is given by 
$$|{\Psi[A]}\rangle = \sum_{p_1,\ldots,p_N}\sum_{a_1,\ldots,a_{N-1}}A^{a_1}_{p_1}A^{a_1,a_2}_{p_2}\ldots A^{a_{N-1}}_{p_N}|p_1,p_2,\ldots,p_N\rangle.$$ 
Here, $p_r$ and $a_r$ label the physical and virtual degrees of freedom at site $r$, respectively. The maximum number of values of $a_r$ is referred to as the bond dimension. In order to represent LSH states using an MPS, the infinite-dimensional Hilbert space associated with the gauge bosons must be truncated. One could either restrict the loop quantum number to $n_l(r)\leq n_{l,\mathrm{max}}$, $\forall r$, leading to a local physical Hilbert-space dimension of $4(n_{l,\mathrm{max}}+1)$. The other choice is to restrict the value of the $N_{L/R}$ quantum number to $N_{L/R}\leq 2J_{\rm max}$, where $J_\text{max}$ is the cutoff on the irreducible representation of the gauge group in the Kogut-Susskind formulation. The truncated LSH state in either case can be represented with an MPS, with $|p_r,\ldots,p_N\rangle \equiv \bigotimes_{r=1 }^N|p_r\rangle$ and $|p_r\rangle \equiv |n_l(r),n_i(r),n_o(r)\rangle$. 

The LSH Hamiltonian admits two $U(1)$ global symmetries resulting in the conservation of charges $Q=\sum_{r=1}^N \big[n_i(r)+n_o(r)\big]$ and $q=\sum_{r=1}^N \big[n_o(r)-n_i(r)\big]$. For OBCs, $Q \in [0,2N]$ and $q \in [-N,N]$, and only certain combinations of $Q$ and $q$ are consistent with the AGL. We restrict ourselves to the $q=0$ and $Q=N$ sector. These global symmetries endow the local tensors with a block-diagonal structure, simplifying algorithmic complexity. Additionally, we impose a penalty term in the Hamiltonian to ensure the satisfaction of the AGL. The LSH Hamiltonian comprises at most nearest-neighbor interactions, allowing a compact matrix-product-operator (MPO) representation. 

Finally, tensor-network techniques rely upon truncating the bond dimension at some maximum value, denoted as $D_{\rm max}$, to enable efficient computations. The tensors $A$ constitute the variational degrees of freedom which are optimized to approximate ground states and perform time evolution. We rely upon existing methods defined in the ITensors.jl package \cite{itensor} to construct and optimize the MPS and MPOs needed for this study.

\section{Probing static properties
\label{sec:static_properties}}
Using standard DMRG methods, the ground state for the dimensionless Hamiltonian 
\begin{equation}
\label{eq:dimless_Ham}
    \hat H \coloneq \frac{2}{ag^2} \hat{H}^{\text{LSH}}=\hat H_E + \mu \hat H_M + x \hat H_I,
\end{equation}
with $ \mu=2\frac{m}{g}\sqrt{x}$ and $ x=\frac{1}{a^2g^2}$, is estimated for two cases: (a) the zero static-charge sector and (b) a pair of spatially separated static charges. The AGL is imposed as a penalty term in the Hamiltonian as $\hat{H}_\text{P} = \Lambda_\text{P}\sum_r \Big[\hat{N}_L(r)-\hat{N}_R(r+1)+\mathcal{Q}_r\Big]^2$ where $\mathcal{Q}_r \in \mathbb{Z}$ originates from the static charges inserted on the lattice.\footnote{This approach is similar to the one followed in Ref.~\cite{kuhn2015non}---it neglects any modifications to the Hamiltonian resulting from the non-vanishing static charges.} We take $\Lambda_\text{P}$ to be proportional 
to the upper bound on the single-site energy, i.e., $\Lambda_\text{P}=2\mu+2x+J_{\rm max}(2J_{\rm max}+1)$, and verify that the computations are effectively restricted to the correct AGL sector.
\setlength{\textfloatsep}{0pt plus 16.0pt minus 2.0pt}
\begin{figure}[h!]
    \centering
    \includegraphics[width=\textwidth]{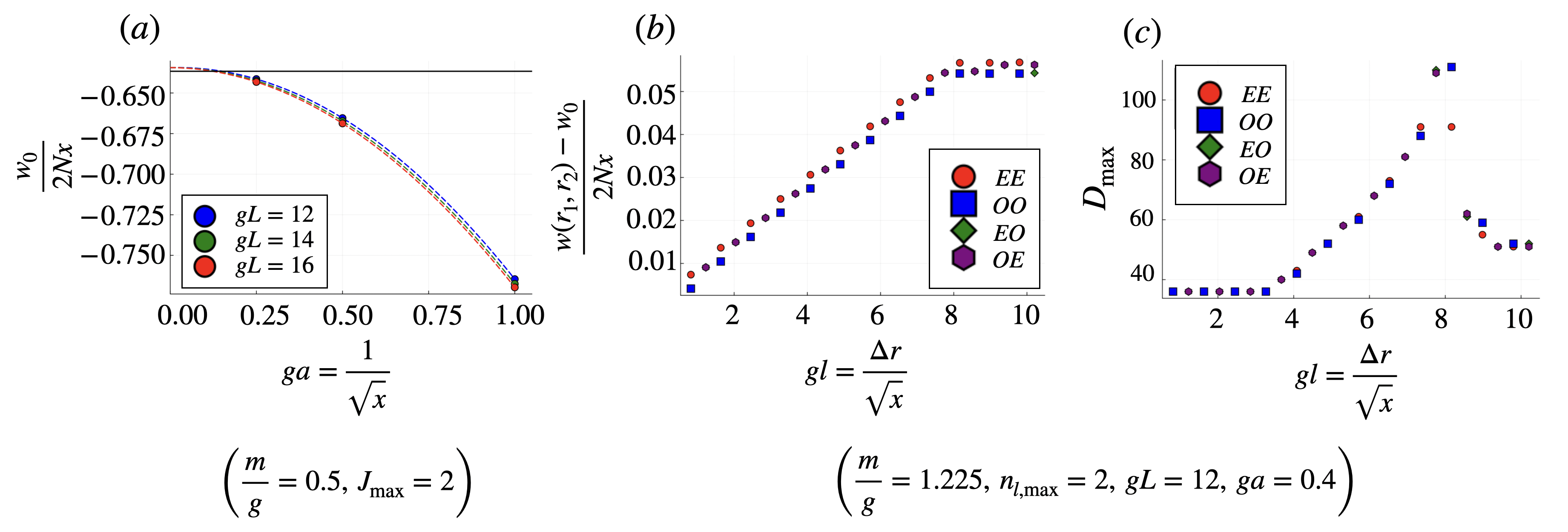}
    \caption{(a) Ground-state energy in the vacuum sector plotted as a function of lattice spacing $ga$ for various system volumes $gL$ and for $J_{\rm max}=2$. The theoretical value (solid black line) corresponds to $\frac{m}{g}=0$ while the DMRG data points correspond to $\frac{m}{g}=0.5$. (b) and (c) Properties of the static-string ground state for $gL=12$, $x=6$, $\frac{m}{g}=1.225$, and a maximum loop quantum number $n_{l,\text{max}}=1$. (b) shows the static potential per unit volume while (c) shows the maximum bond dimension as a function of the static-string length $gl$. $EE,\ OO,\ EO$, and $OE$ refer to choices of placements of the string endpoints at even (E) or odd (O) sites. The values corresponding to $OE$ and $EO$ strings often coincide, thus the green diamonds are often hidden behind the purple hexagons.    }
    \label{fig:static_results}
\end{figure}

We first consider the zero static-charge sector, i.e., $\mathcal{Q}_r=0$ $\forall r$. The initial state is taken to be the strong-coupling vacuum defined as fully filled odd sites $|n_l=0,n_i=1,n_o=1\rangle$ and empty even sites $|n_l=0,n_i=0,n_o=0\rangle$.
The dimensionless ground-state energy is denoted as $w_0=\frac{2}{ag^2}E_0$, where $E_0$ is the ground-state energy associated with the dimensionful Hamiltonian $\hat H$.

The lattice computations are specified by dimensionless quantities $N,~x,~\frac{m}{g}$, and $J_\text{max}$ or $n_{l,\mathrm{max}}$, corresponding to a lattice with (dimensionless) lattice spacing $ga=\frac{1}{\sqrt{x}}$ and (dimensionless) volume $gL=\frac{N}{\sqrt{x}}$ with $L \coloneq Na$. The continuum limit corresponds to the following ordered limit at fixed $\frac{m}{g}$: At a fixed $x$, take $J_\text{max}$ or $n_{l,\mathrm{max}}\to \infty$ followed by $N \to \ \infty$. Finally, take $x \to \infty$~\cite{Hamer:1981yq}.

We compute the value of the ground-state energy for $\frac{m}{g}=0.5$ and $J_{\rm max}=2$ for various values of $ga$ and system volumes $gL$ in Fig.~\ref{fig:static_results}(a). The solid black line corresponds to the analytical continuum value of $-\frac{2}{\pi}$ for $\frac{m}{g}=0$~\cite{hamer19822,banuls2017efficient}. A complete continuum-limit study of this energy is in progress (the dotted line is to guide the eye and is not yet a fit to data). 

Next, we compute the ground-state energy in the external static-charge sector: we insert $\mathcal{Q}_{r_1} = -1$ and $\mathcal{Q}_{r_2}=+1$ at sites $r_1$ and $r_2$, respectively, with $r_1<r_2$, and leave $\mathcal{Q}_r = 0$ for $r\neq r_1,\,r_2$. 
The static potential is obtained by subtracting the ground-state energy in the zero static-charge sector, $w_0$, from the ground state energy $w(r_1,r_2)$ for different string lengths $gl \coloneq ga(r_2-r_1) \equiv ga \Delta r$, where both $w_0$ and  $w(r_1,r_2)$ are calculated via DMRG. Figure~\ref{fig:static_results}(b) represents the static potential per unit volume as a function of the static-string length for different choices of static-charge placement: the string begins/ends at an even (E) site or odd (O) site, and for $gL=12$, $ga=0.4$, and $\frac{m}{g}=1.225$. These plots are generated by imposing a maximum loop-quantum-number cutoff of $n_{l,\text{max}}=1$. A small difference in the string potential is observed, but the overall behavior is consistent across all four choices of string endpoints. The linear part of the potential corresponds to the unbroken string while the plateau region signals the broken string.

Figure~\ref{fig:static_results}(c) shows the maximum bond dimension the MPS requires to reach convergence as a function of $gl$. We set the DMRG's convergence threshold to less than $10^{-8}$. The maximum bond dimension peaks around the length of $gl\sim 8$, at which the string breaks, indicating a larger amount of entanglement in the state at the transition point.

While not shown here, we have also performed computations for the static potential for a larger volume $gL=16$ and a smaller lattice spacing $ga=0.25$ at $J_\text{max}=2$ at different values of $\frac{m}{g}$. We observe that as the quark mass is decreased, the amount of entanglement in the system grows, consistent with a faster approach to the string-breaking point. The complete result of this analysis will be reported elsewhere. These results are in qualitative agreement with those in Ref.~\cite{kuhn2015non} which employed a smaller cutoff.

\section{Dynamic properties}
\label{sec:dynamic_properties}
With the MPS ansatz described in Sec.~\ref{sec:LSH_network}, one can set up a dynamical study of this model as well. In particular, we aim to study the string-breaking phenomenon for an initial (bare) mesonic state. This state is constructed from the interacting vacuum by applying a ``string'' operator, i.e, a gauge-invariant-antiquark bilinear operator where the pair is separated by distance $gl=ga\Delta r$. This state, which is not an eigenstate of the LHS Hamiltonian, is then evolved under the LSH Hamiltonian. The extended meson is expected to disintegrate into shorter mesons due to confinement.

In the LSH formulation, this (bare) mesonic state can be written as:
\begin{equation}
\label{eq:String_state_I}
    |{S[A]}\rangle = \hat{S}_{r,\Delta r}|{\Omega[A]}\rangle,
\end{equation}
where we choose the string operator to be $\hat{S}_{r,\Delta r} \coloneq \frac{1}{ga}\Big(\hat{S}^\text{LSH}_{r,\Delta r}-\hat{S}^\text{LSH}_{r+1,\Delta r}\Big)$.
Here, $r$ is assumed odd and 
$\Delta r$ is assumed even. This choice in the staggered formulation turns out to be consistent with the string operator in the continuum formulation. $|{\Omega[A]}\rangle$ in Eq.~\eqref{eq:String_state_I} is the ground state of the 
Hamiltonian defined in Eq.~\eqref{eq:dimless_Ham} obtained via DMRG. The string operator of the LSH formulation can be derived from the string operator of the Kogut-Susskind formulation (i.e., $\hat{{\Psi}}^\dagger_r \hat{U}_{r+1}\hat{U}_{r+2}\ldots \hat{U}_{r+\Delta r-1}\hat{\Psi}_{r+\Delta r}$ with $\Psi$ being the staggered fermion operator and $U$ being the gauge-link operator) using the mapping between the LSH and Kogut-Susskind operators. The result is:
\begin{align}
\label{String_operator_LSH}
    \hat{S}^\text{LSH}_{r,\Delta r} 
=\sum_{\sigma_1,\sigma_2,\ldots,\sigma_{\Delta r}=\pm}
    \frac{1}{\sqrt{\hat N_L(r)+1}}\hat{\mathcal{S}}_{\text{out}}^{+,\sigma_1}(r)\hat{\bar{\mathcal{L}}}^{\sigma_1,\sigma_2}
    &(r+1)\ldots  \hat{\bar{\mathcal{L}}}^{\sigma_{\Delta r-1},\sigma_{\Delta r}}
    (r+\Delta r-1)
      \nonumber\\
    &\times \hat{\mathcal{S}}^{\sigma_{\Delta r},-}_{\text{in}}(r+\Delta r)
    \frac{1}{\sqrt{\hat N_R(r+\Delta r)+1}},
\end{align}
where the string ($\mathcal{\hat S}^{\sigma_1,\sigma_2}_{\text{out/in}}$), loop ($\hat{\bar{\mathcal{L}}}^{\sigma_1,\sigma_2}$), and diagonal $\hat N_{L/R}$ operators are defined in Table~\ref{tab:lsh_dict}.
\begin{figure*}
    \centering
    \includegraphics[width=0.9\textwidth]{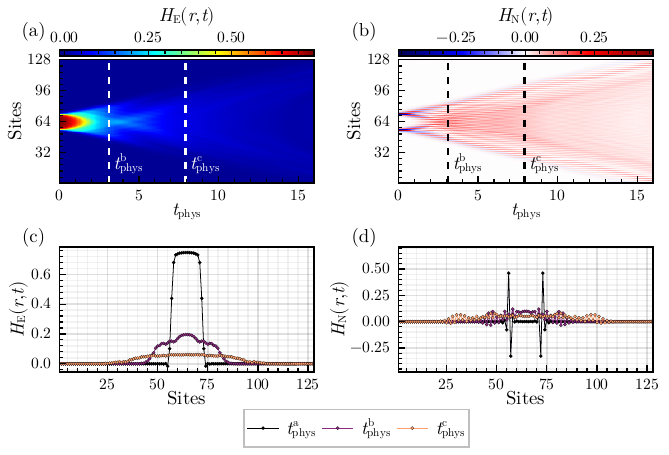}
    \caption{(a) Ground-state-subtracted site-local electric-field density, Eq.~\eqref{eq:HE_r}, and (b) fermion-number density, Eq.~\eqref{eq:H_stagg}, as a function of time. (c) and (d) are the same quantities as in (a) and (b) but at single time slices $t^\text{a}_\text{phys}=0.0,~t^\text{b}_\text{phys}=3.12$, and $ t^\text{c}_\text{phys}=7.92$. (Note that $t_\text{a}$ coincides with the   y-axis in the top subplots.)
    }
    \label{fig:dynamic}
\end{figure*}

Once the initial mesonic state is evolved for time $t$, it turns into state $|\psi(t) \rangle \coloneq e^{-it\hat H} |S[A]\rangle$. Here, $\hat H$ is the dimensionless Hamiltonian defined in Eq.~\eqref{eq:dimless_Ham}, and $t$ is dimensionless time. The following set of observables are then used to analyze the dynamics of string breaking:
\begin{enumerate}
    \item (Ground-state subtracted) instantaneous electric-flux density
     at each lattice site,
    \begin{equation}
    \label{eq:HE_r}
        H_\text{E}(r,t) \coloneq 
        \langle \psi(t) |\hat h_\text{E}(r) |\psi(t) \rangle - \langle \Omega[A] |\hat h_\text{E}(r) |\Omega[A] \rangle,
    \end{equation}
    where $\hat h_\text{E}(r) \coloneq \frac{\hat{N}_L(r)}{4}
        \Big(\frac{\hat{N}_L(r)}{2}+1\Big)+ \frac{\hat{N}_R(r)}{4}
        \Big(\frac{\hat{N}_R(r)}{2}+1\Big)$.
    \item (Ground-state-subtracted) instantaneous fermion-number density at each lattice site,
\begin{equation}
    \label{eq:H_stagg}
    H_\text{N}(r,t) \coloneq 
    \langle \psi(t) |[\hat{n}_i(r)+\hat{n}_o(r)]|\psi(t) \rangle 
    - \langle \Omega[A] |[\hat{n}_i(r)+\hat{n}_o(r)]|\Omega[A] \rangle .
    \end{equation}
    \item Loschmidt-echo rate function,
    \begin{equation}
    \label{eq:Loschmidt}
    \lambda(t) = - 
    \lim _{N \to \infty} 
    \frac{1}{N} 
    \log(|\mathcal{G}(t)|),
    \end{equation}
    where $\mathcal{G}(t) \coloneq \langle S[A]|
    \psi(t) \rangle$. This rate is used to detect the transition from a longer initial string to a collection of shorter strings with small to vanishing overlap to the initial state.
\end{enumerate}

We employ the 2-site time-dependent variational-principle (TDVP) algorithm~\cite{Haegeman:2011zz,Haegeman:2016gfj} to evolve the initial mesonic state with the Hamiltonian in Eq.~\eqref{eq:dimless_Ham} for the lattice parameters $N=128,~x=16,
\frac{m}{g}=0.2,
~J_\text{max}=5/2,~D_\text{max}=200$, $T=2$, and $dt=0.01$. $D_\text{max}$ is the maximum bond dimension that is dynamically set by the TDVP algorithm, $T$ is the total (dimensionless) simulation time, and $dt$ is the chosen TDVP time step. The dimensionful time is defined as $t_\text{phys}=\big(\frac{2}{ag^2}\big) t$. Figure~\ref{fig:dynamic}(a) displays the heatmap of $H_\text{E}$ in the $(t_\text{phys},r)$ plane. The initial string of length $gl=16$, positioned at the center of the lattice, begins to break apart from its ends as time progresses. This observation aligns with the heatmap of the number-density displayed in Fig.~\ref{fig:dynamic}(b). Streams of particle and antiparticle pairs can be seen moving both inward and outward, resulting in subsequent collision processes and the generation of several new pairs.
\setlength{\textfloatsep}{0pt plus 24.0pt minus 2.0pt}
\begin{figure*}
    \includegraphics[width=\textwidth]{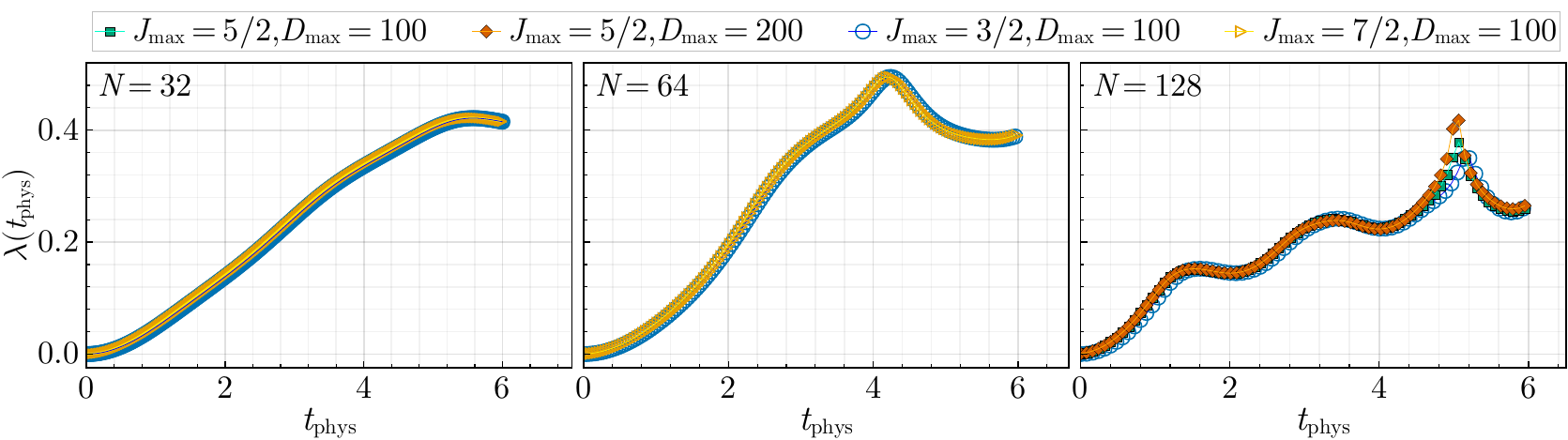}
    \caption{Loschmidt echo plotted as a function of $t_\text{phys}$ for fixed $\frac{m}{g}=0.2$ parameter sets (a) $\{N,x,\Delta r,T\}=\{32,1,4,3\}$, (b) $\{64,4,8,1.5\}$, and (c) $\{128,16,16,0.75\}$.}
    \label{fig:echo}
\end{figure*}

Figures~\ref{fig:dynamic}(c) and (d) represent the local electric-field and fermion-number density at the specified time slices $t^{\text{a/b/c}}_\text{phys}$, respectively. These plots demonstrate the change in the distribution of electric-energy density from localized (string-like) in the center toward delocalized outward. This result aligns with the abundant generation of particles in later times. These plots display richer phenomenology compared with existing work~\cite{kuhn2015non,sala2018variational}, given the larger Hilbert-space cutoff and longer dynamical strings employed in this work. A comprehensive analysis of these dynamics, including entanglement generation, correlations, and error estimates will be reported in an upcoming publication.

Finally, the Loschmidt-echo rate function $\lambda$ is used to probe the change in the initial state. Non-analyticities in the rate function correspond to times of least overlap between the initial and time-evolved states, thus signifying string breaking. We probe increasingly finer lattices at a fixed bare mass $\frac{m}{g}=0.2$, physical volume $gL=32$, and physical string length $ga\Delta r=4$ in Fig.~\ref{fig:echo}. The Loschmidt-echo profiles line up closely as $J_{\rm max}$ and $D_{\rm max}$ are varied for $x=1,4$. More pronounced bond-dimension truncation effects are observed for the finest lattice spacing of $x=16$ at later times. However, the profiles fluctuate significantly as the lattice spacing is decreased. This suggests a need to use even finer lattice spacings to observe a stable profile and obtain the critical time associated with the peak of the Loschmidt echo in the $a\rightarrow 0$ limit. Such a calculation requires a simultaneous increase in the bond dimension to ensure accuracy. Work is in progress in this direction. 

\section{Conclusion}
\label{sec:conclusion}
This work introduces a tensor-network framework within the loop-string-hadron formulation of the SU(2) lattice gauge theory in (1+1)D. We compute the ground-state energy in this theory in the zero and non-zero static-charge sectors. Additionally, we perform simulations of dynamical string breaking that yield rich phenomenology. Our results extend the reach of previous work to larger systems, larger Hilbert-space truncation cutoffs, and longer strings, with the possibility of continuum-limit extrapolation even for dynamical quantities. A complete analysis of these properties, including continuum-limit extrapolations, is underway.

A key takeaway from this work is that the LSH framework is a suitable starting point for tensor-network studies of non-Abelian gauge theories: it only involves gauge-invariant degrees of freedom, retains the locality of Hamiltonian, requires the imposition of only Abelian constraints, and is generalized to any dimension and other boundary conditions.
The simplified construction of dynamical fermions will facilitate creating interacting wave packets in scattering simulations, and the Abelianized form will likely enable efficient tensor-network ansatze in higher dimensions. Generalization of these calculations to the SU(3) LGT is underway.

\acknowledgments
We acknowledge Niklas Mueller's early contributions to this work. This work was supported by fellowship support from Birla Institute of Technology and Science (BITS)-Pilani and the International Travel Support (ITS/2024/002694) from ANRF, India; the U.S. National Science Foundation's Quantum Leap Challenge Institute (OMA-2120757); Maryland Center for Fundamental Physics, Department of Physics, and College of Computer, Mathematical, and Natural Sciences at the University of Maryland; the U.S. Department of Energy, Office of Science (InQubator for Quantum Simulation with grant no. DE-SC0020970 via the NP Quantum Horizons program, the HEP QuantISED program with Fermilab subcontract no. 666484 and the HEP QuantISED program KA2401032 under contract no. DE-AC02-05CH11231, Early Career Award DE-SC0020271, and the ASCR Far-Qu project); the Department of Physics and the College of Arts and Sciences at the University of Washington; the  OPERA award (FR/SCM/11-Dec-2020/PHY) from BITS-Pilani, the Start-up Research Grant (SRG/2022/000972) and Core-Research Grant (CRG/2022/007312) from ANRF, India, and the cross-discipline research fund (C1/23/185) from BITS-Pilani. We acknowledge computing resources from the Zaratan HPC cluster at the University of Maryland, a BITS Pilani, Goa cluster and the Sharanga HPC cluster at the BITS-Pilani, Hyderabad. 

\bibliographystyle{JHEP}
\bibliography{bibi.bib}

\providecommand{\href}[2]{#2}\begingroup\raggedright\begin{thebibliography}{10}

\bibitem{white1992density}
S.R.~White, \emph{{Density matrix formulation for quantum renormalization
  groups}}, \href{https://doi.org/10.1103/PhysRevLett.69.2863}{\emph{Phys. Rev.
  Lett.} {\bfseries 69} (1992) 2863}.

\bibitem{fannes1992finitely}
M.~Fannes, B.~Nachtergaele and R.F.~Werner, \emph{{Finitely Correlated States
  on Quantum Spin Chains}},
  \href{https://doi.org/10.1007/BF02099178}{\emph{Commun. Math. Phys.}
  {\bfseries 144} (1992) 443}.

\bibitem{ostlund1995thermodynamic}
S.~Ostlund and S.~Rommer, \emph{{Thermodynamic Limit of Density Matrix
  Renormalization for the spin-1 heisenberg chain}},
  \href{https://doi.org/10.1103/PhysRevLett.75.3537}{\emph{Phys. Rev. Lett.}
  {\bfseries 75} (1995) 3537}
  [\href{https://arxiv.org/abs/cond-mat/9503107}{{\ttfamily
  cond-mat/9503107}}].

\bibitem{verstraete2004density}
F.~Verstraete, D.~Porras and J.I.~Cirac, \emph{{Density Matrix Renormalization
  Group and Periodic Boundary Conditions: A Quantum Information Perspective}},
  \href{https://doi.org/10.1103/PhysRevLett.93.227205}{\emph{Phys. Rev. Lett.}
  {\bfseries 93} (2004) 227205}
  [\href{https://arxiv.org/abs/cond-mat/0404706}{{\ttfamily
  cond-mat/0404706}}].

\bibitem{schollwock2011density}
U.~Schollwoeck, \emph{{The density-matrix renormalization group in the age of
  matrix product states}},
  \href{https://doi.org/10.1016/j.aop.2010.09.012}{\emph{Annals Phys.}
  {\bfseries 326} (2011) 96} [\href{https://arxiv.org/abs/1008.3477}{{\ttfamily
  1008.3477}}].

\bibitem{cirac2020matrix}
J.I.~Cirac, D.~Perez-Garcia, N.~Schuch and F.~Verstraete, \emph{{Matrix product
  states and projected entangled pair states: Concepts, symmetries, theorems}},
  \href{https://doi.org/10.1103/RevModPhys.93.045003}{\emph{Rev. Mod. Phys.}
  {\bfseries 93} (2021) 045003}
  [\href{https://arxiv.org/abs/2011.12127}{{\ttfamily 2011.12127}}].

\bibitem{meurice2020tensor}
Y.~Meurice, R.~Sakai and J.~Unmuth-Yockey, \emph{{Tensor lattice field theory
  for renormalization and quantum computing}},
  \href{https://doi.org/10.1103/RevModPhys.94.025005}{\emph{Rev. Mod. Phys.}
  {\bfseries 94} (2022) 025005}
  [\href{https://arxiv.org/abs/2010.06539}{{\ttfamily 2010.06539}}].

\bibitem{banuls2020review}
M.C.~Ba\~nuls and K.~Cichy, \emph{{Review on Novel Methods for Lattice Gauge
  Theories}}, \href{https://doi.org/10.1088/1361-6633/ab6311}{\emph{Rept. Prog.
  Phys.} {\bfseries 83} (2020) 024401}
  [\href{https://arxiv.org/abs/1910.00257}{{\ttfamily 1910.00257}}].

\bibitem{byrnes2002density}
T.~Byrnes, P.~Sriganesh, R.J.~Bursill and C.J.~Hamer, \emph{{Density matrix
  renormalization group approach to the massive Schwinger model}},
  \href{https://doi.org/10.1016/S0920-5632(02)01416-0}{\emph{Nucl. Phys. B
  Proc. Suppl.} {\bfseries 109} (2002) 202}
  [\href{https://arxiv.org/abs/hep-lat/0201007}{{\ttfamily hep-lat/0201007}}].

\bibitem{tagliacozzo2011entanglement}
L.~Tagliacozzo and G.~Vidal, \emph{{Entanglement Renormalization and Gauge
  Symmetry}}, \href{https://doi.org/10.1103/PhysRevB.83.115127}{\emph{Phys.
  Rev. B} {\bfseries 83} (2011) 115127}
  [\href{https://arxiv.org/abs/1007.4145}{{\ttfamily 1007.4145}}].

\bibitem{rico2014tensor}
E.~Rico, T.~Pichler, M.~Dalmonte, P.~Zoller and S.~Montangero, \emph{{Tensor
  networks for Lattice Gauge Theories and Atomic Quantum Simulation}},
  \href{https://doi.org/10.1103/PhysRevLett.112.201601}{\emph{Phys. Rev. Lett.}
  {\bfseries 112} (2014) 201601}
  [\href{https://arxiv.org/abs/1312.3127}{{\ttfamily 1312.3127}}].

\bibitem{buyens2014matrix}
B.~Buyens, J.~Haegeman, K.~Van~Acoleyen, H.~Verschelde and F.~Verstraete,
  \emph{{Matrix product states for gauge field theories}},
  \href{https://doi.org/10.1103/PhysRevLett.113.091601}{\emph{Phys. Rev. Lett.}
  {\bfseries 113} (2014) 091601}
  [\href{https://arxiv.org/abs/1312.6654}{{\ttfamily 1312.6654}}].

\bibitem{silvi2014lattice}
P.~Silvi, E.~Rico, T.~Calarco and S.~Montangero, \emph{{Lattice Gauge Tensor
  Networks}}, \href{https://doi.org/10.1088/1367-2630/16/10/103015}{\emph{New
  J. Phys.} {\bfseries 16} (2014) 103015}
  [\href{https://arxiv.org/abs/1404.7439}{{\ttfamily 1404.7439}}].

\bibitem{tagliacozzo2014tensor}
L.~Tagliacozzo, A.~Celi and M.~Lewenstein, \emph{{Tensor Networks for Lattice
  Gauge Theories with continuous groups}},
  \href{https://doi.org/10.1103/PhysRevX.4.041024}{\emph{Phys. Rev. X}
  {\bfseries 4} (2014) 041024}
  [\href{https://arxiv.org/abs/1405.4811}{{\ttfamily 1405.4811}}].

\bibitem{zohar2015formulation}
E.~Zohar and M.~Burrello, \emph{{Formulation of lattice gauge theories for
  quantum simulations}},
  \href{https://doi.org/10.1103/PhysRevD.91.054506}{\emph{Phys. Rev. D}
  {\bfseries 91} (2015) 054506}
  [\href{https://arxiv.org/abs/1409.3085}{{\ttfamily 1409.3085}}].

\bibitem{rigobello2021entanglement}
M.~Rigobello, S.~Notarnicola, G.~Magnifico and S.~Montangero,
  \emph{{Entanglement generation in (1+1)D QED scattering processes}},
  \href{https://doi.org/10.1103/PhysRevD.104.114501}{\emph{Phys. Rev. D}
  {\bfseries 104} (2021) 114501}
  [\href{https://arxiv.org/abs/2105.03445}{{\ttfamily 2105.03445}}].

\bibitem{frias2022light}
M.~Fr\'\i{}as-P\'erez and M.C.~Ba\~nuls, \emph{{Light cone tensor network and
  time evolution}},
  \href{https://doi.org/10.1103/PhysRevB.106.115117}{\emph{Phys. Rev. B}
  {\bfseries 106} (2022) 115117}
  [\href{https://arxiv.org/abs/2201.08402}{{\ttfamily 2201.08402}}].

\bibitem{canals2024tensor}
M.~Canals, N.~Chepiga and L.~Tagliacozzo, \emph{{A tensor network formulation
  of Lattice Gauge Theories based only on symmetric tensors}},
  \href{https://arxiv.org/abs/2412.16961}{{\ttfamily 2412.16961}}.

\bibitem{banuls2023quantum}
M.C.~Banuls, M.P.~Heller, K.~Jansen, J.~Knaute and V.~Svensson, \emph{{Quantum
  information perspective on meson melting}},
  \href{https://doi.org/10.1103/PhysRevD.108.076016}{\emph{Phys. Rev. D}
  {\bfseries 108} (2023) 076016}
  [\href{https://arxiv.org/abs/2206.10528}{{\ttfamily 2206.10528}}].

\bibitem{funcke2023exploring}
L.~Funcke, K.~Jansen and S.~K\"uhn, \emph{{Exploring the CP-violating Dashen
  phase in the Schwinger model with tensor networks}},
  \href{https://doi.org/10.1103/PhysRevD.108.014504}{\emph{Phys. Rev. D}
  {\bfseries 108} (2023) 014504}
  [\href{https://arxiv.org/abs/2303.03799}{{\ttfamily 2303.03799}}].

\bibitem{magnifico2024tensor}
G.~Magnifico, G.~Cataldi, M.~Rigobello, P.~Majcen, D.~Jaschke, P.~Silvi et~al.,
  \emph{{Tensor Networks for Lattice Gauge Theories beyond one dimension: a
  Roadmap}},  \href{https://arxiv.org/abs/2407.03058}{{\ttfamily 2407.03058}}.

\bibitem{banuls2024parton}
M.~Schneider, M.C.~Ba\~nuls, K.~Cichy and C.J.D.~Lin, \emph{{Parton
  Distribution Functions in the Schwinger Model with Tensor Networks}},
  \href{https://doi.org/10.22323/1.466.0024}{\emph{PoS} {\bfseries LATTICE2024}
  (2025) 024} [\href{https://arxiv.org/abs/2409.16996}{{\ttfamily
  2409.16996}}].

\bibitem{belyansky2024high}
R.~Belyansky, S.~Whitsitt, N.~Mueller, A.~Fahimniya, E.R.~Bennewitz, Z.~Davoudi
  et~al., \emph{{High-Energy Collision of Quarks and Mesons in the Schwinger
  Model: From Tensor Networks to Circuit QED}},
  \href{https://doi.org/10.1103/PhysRevLett.132.091903}{\emph{Phys. Rev. Lett.}
  {\bfseries 132} (2024) 091903}
  [\href{https://arxiv.org/abs/2307.02522}{{\ttfamily 2307.02522}}].

\bibitem{kuhn2015non}
S.~K\"uhn, E.~Zohar, J.I.~Cirac and M.C.~Ba\~nuls, \emph{{Non-Abelian string
  breaking phenomena with Matrix Product States}},
  \href{https://doi.org/10.1007/JHEP07(2015)130}{\emph{JHEP} {\bfseries 07}
  (2015) 130} [\href{https://arxiv.org/abs/1505.04441}{{\ttfamily
  1505.04441}}].

\bibitem{banuls2017efficient}
M.C.~Ba\~nuls, K.~Cichy, J.I.~Cirac, K.~Jansen and S.~K\"uhn, \emph{{Efficient
  basis formulation for 1+1 dimensional SU(2) lattice gauge theory: Spectral
  calculations with matrix product states}},
  \href{https://doi.org/10.1103/PhysRevX.7.041046}{\emph{Phys. Rev. X}
  {\bfseries 7} (2017) 041046}
  [\href{https://arxiv.org/abs/1707.06434}{{\ttfamily 1707.06434}}].

\bibitem{sala2018variational}
P.~Sala, T.~Shi, S.~K\"uhn, M.C.~Ba\~nuls, E.~Demler and J.I.~Cirac,
  \emph{{Variational study of U(1) and SU(2) lattice gauge theories with
  Gaussian states in 1+1 dimensions}},
  \href{https://doi.org/10.1103/PhysRevD.98.034505}{\emph{Phys. Rev. D}
  {\bfseries 98} (2018) 034505}
  [\href{https://arxiv.org/abs/1805.05190}{{\ttfamily 1805.05190}}].

\bibitem{silvi2019tensor}
P.~Silvi, Y.~Sauer, F.~Tschirsich and S.~Montangero, \emph{{Tensor network
  simulation of an SU(3) lattice gauge theory in 1D}},
  \href{https://doi.org/10.1103/PhysRevD.100.074512}{\emph{Phys. Rev. D}
  {\bfseries 100} (2019) 074512}
  [\href{https://arxiv.org/abs/1901.04403}{{\ttfamily 1901.04403}}].

\bibitem{rigobello2023hadrons}
M.~Rigobello, G.~Magnifico, P.~Silvi and S.~Montangero, \emph{{Hadrons in
  (1+1)D Hamiltonian hardcore lattice QCD}},
  \href{https://arxiv.org/abs/2308.04488}{{\ttfamily 2308.04488}}.

\bibitem{hayata2024dense}
T.~Hayata, Y.~Hidaka and K.~Nishimura, \emph{{Dense QCD$_{2}$ with matrix
  product states}}, \href{https://doi.org/10.1007/JHEP07(2024)106}{\emph{JHEP}
  {\bfseries 07} (2024) 106}
  [\href{https://arxiv.org/abs/2311.11643}{{\ttfamily 2311.11643}}].

\bibitem{Bender:2020jgr}
J.~Bender, P.~Emonts, E.~Zohar and J.I.~Cirac, \emph{{Real-time dynamics in
  2+1d compact QED using complex periodic Gaussian states}},
  \href{https://doi.org/10.1103/PhysRevResearch.2.043145}{\emph{Phys. Rev.
  Res.} {\bfseries 2} (2020) 043145}
  [\href{https://arxiv.org/abs/2006.10038}{{\ttfamily 2006.10038}}].

\bibitem{emonts2020variational}
P.~Emonts, M.C.~Ba\~nuls, I.~Cirac and E.~Zohar, \emph{{Variational Monte Carlo
  simulation with tensor networks of a pure $\mathbb{Z}_3$ gauge theory in
  (2+1)d}}, \href{https://doi.org/10.1103/PhysRevD.102.074501}{\emph{Phys. Rev.
  D} {\bfseries 102} (2020) 074501}
  [\href{https://arxiv.org/abs/2008.00882}{{\ttfamily 2008.00882}}].

\bibitem{magnifico2021lattice}
G.~Magnifico, T.~Felser, P.~Silvi and S.~Montangero, \emph{{Lattice quantum
  electrodynamics in (3+1)-dimensions at finite density with tensor networks}},
  \href{https://doi.org/10.1038/s41467-021-23646-3}{\emph{Nature Commun.}
  {\bfseries 12} (2021) 3600}
  [\href{https://arxiv.org/abs/2011.10658}{{\ttfamily 2011.10658}}].

\bibitem{robaina2021simulating}
D.~Robaina, M.C.~Ba\~nuls and J.I.~Cirac, \emph{{Simulating $2+1D$ $Z_3$
  Lattice Gauge Theory with an Infinite Projected Entangled-Pair State}},
  \href{https://doi.org/10.1103/PhysRevLett.126.050401}{\emph{Phys. Rev. Lett.}
  {\bfseries 126} (2021) 050401}
  [\href{https://arxiv.org/abs/2007.11630}{{\ttfamily 2007.11630}}].

\bibitem{Emonts:2022yom}
P.~Emonts, A.~Kelman, U.~Borla, S.~Moroz, S.~Gazit and E.~Zohar, \emph{{Finding
  the ground state of a lattice gauge theory with fermionic tensor networks: A
  $(2+1)D$ $Z_2$ demonstration}},
  \href{https://doi.org/10.1103/PhysRevD.107.014505}{\emph{Phys. Rev. D}
  {\bfseries 107} (2023) 014505}
  [\href{https://arxiv.org/abs/2211.00023}{{\ttfamily 2211.00023}}].

\bibitem{Cataldi:2023xki}
G.~Cataldi, G.~Magnifico, P.~Silvi and S.~Montangero, \emph{{Simulating
  $(2+1)D$ SU(2) Yang-Mills lattice gauge theory at finite density with tensor
  networks}},
  \href{https://doi.org/10.1103/PhysRevResearch.6.033057}{\emph{Phys. Rev.
  Res.} {\bfseries 6} (2024) 033057}
  [\href{https://arxiv.org/abs/2307.09396}{{\ttfamily 2307.09396}}].

\bibitem{Kelman:2024exo}
A.~Kelman, U.~Borla, P.~Emonts and E.~Zohar, \emph{{Projected Entangled Pair
  States for Lattice Gauge Theories with Dynamical Fermions}},
  \href{https://arxiv.org/abs/2412.16951}{{\ttfamily 2412.16951}}.

\bibitem{banuls2018tensor}
M.C.~Ba\~nuls, K.~Cichy, J.I.~Cirac, K.~Jansen and S.~K\"uhn, \emph{{Tensor
  Networks and their use for Lattice Gauge Theories}},
  \href{https://doi.org/10.22323/1.334.0022}{\emph{PoS} {\bfseries LATTICE2018}
  (2018) 022} [\href{https://arxiv.org/abs/1810.12838}{{\ttfamily
  1810.12838}}].

\bibitem{meurice2022tensor}
Y.~Meurice, R.~Sakai and J.~Unmuth-Yockey, \emph{{Tensor lattice field theory
  for renormalization and quantum computing}},
  \href{https://doi.org/10.1103/RevModPhys.94.025005}{\emph{Rev. Mod. Phys.}
  {\bfseries 94} (2022) 025005}
  [\href{https://arxiv.org/abs/2010.06539}{{\ttfamily 2010.06539}}].

\bibitem{Magnifico:2024eiy}
G.~Magnifico, G.~Cataldi, M.~Rigobello, P.~Majcen, D.~Jaschke, P.~Silvi et~al.,
  \emph{{Tensor Networks for Lattice Gauge Theories beyond one dimension: a
  Roadmap}},  \href{https://arxiv.org/abs/2407.03058}{{\ttfamily 2407.03058}}.

\bibitem{Raychowdhury:2019iki}
I.~Raychowdhury and J.R.~Stryker, \emph{{Loop, String, and Hadron Dynamics in
  SU(2) Hamiltonian Lattice Gauge Theories}},
  \href{https://doi.org/10.1103/PhysRevD.101.114502}{\emph{Phys. Rev. D}
  {\bfseries 101} (2020) 114502}
  [\href{https://arxiv.org/abs/1912.06133}{{\ttfamily 1912.06133}}].

\bibitem{davoudi2021search}
Z.~Davoudi, I.~Raychowdhury and A.~Shaw, \emph{{Search for efficient
  formulations for Hamiltonian simulation of non-Abelian lattice gauge
  theories}}, \href{https://doi.org/10.1103/PhysRevD.104.074505}{\emph{Phys.
  Rev. D} {\bfseries 104} (2021) 074505}
  [\href{https://arxiv.org/abs/2009.11802}{{\ttfamily 2009.11802}}].

\bibitem{Cochran:2024rwe}
T.A.~Cochran et~al., \emph{{Visualizing Dynamics of Charges and Strings in
  (2+1)D Lattice Gauge Theories}},
  \href{https://arxiv.org/abs/2409.17142}{{\ttfamily 2409.17142}}.

\bibitem{De:2024smi}
A.~De et~al., \emph{{Observation of string-breaking dynamics in a quantum
  simulator}},  \href{https://arxiv.org/abs/2410.13815}{{\ttfamily
  2410.13815}}.

\bibitem{Gonzalez-Cuadra:2024xul}
D.~Gonzalez-Cuadra et~al., \emph{{Observation of string breaking on a (2 + 1)D
  Rydberg quantum simulator}},
  \href{https://arxiv.org/abs/2410.16558}{{\ttfamily 2410.16558}}.

\bibitem{Ciavarella:2024lsp}
A.N.~Ciavarella, \emph{{String Breaking in the Heavy Quark Limit with Scalable
  Circuits}},  \href{https://arxiv.org/abs/2411.05915}{{\ttfamily 2411.05915}}.

\bibitem{Crippa:2024hso}
A.~Crippa, K.~Jansen and E.~Rinaldi, \emph{{Analysis of the confinement string
  in (2 + 1)-dimensional Quantum Electrodynamics with a trapped-ion quantum
  computer}},  \href{https://arxiv.org/abs/2411.05628}{{\ttfamily 2411.05628}}.

\bibitem{Liu:2024lut}
Y.~Liu, W.-Y.~Zhang, Z.-H.~Zhu, M.-G.~He, Z.-S.~Yuan and J.-W.~Pan,
  \emph{{String breaking mechanism in a lattice Schwinger model simulator}},
  \href{https://arxiv.org/abs/2411.15443}{{\ttfamily 2411.15443}}.

\bibitem{kogut1975hamiltonian}
J.B.~Kogut and L.~Susskind, \emph{{Hamiltonian Formulation of Wilson's Lattice
  Gauge Theories}}, \href{https://doi.org/10.1103/PhysRevD.11.395}{\emph{Phys.
  Rev. D} {\bfseries 11} (1975) 395}.

\bibitem{itensor}
M.~Fishman, S.R.~White and E.M.~Stoudenmire, \emph{{The ITensor Software
  Library for Tensor Network Calculations}},
  \href{https://doi.org/10.21468/SciPostPhysCodeb.4}{\emph{SciPost Phys.
  Codebases} (2022) 4} [\href{https://arxiv.org/abs/2007.14822}{{\ttfamily
  2007.14822}}].

\bibitem{Hamer:1981yq}
C.~Hamer, \emph{{SU(2) {Yang-Mills} Theory in (1+1)-dimensions: A Finite
  Lattice Approach}},
  \href{https://doi.org/10.1016/0550-3213(82)90009-8}{\emph{Nucl. Phys. B}
  {\bfseries 195} (1982) 503}.

\bibitem{hamer19822}
C.J.~Hamer, \emph{{SU(2) {Yang-Mills} Theory in (1+1)-dimensions: A Finite
  Lattice Approach}},
  \href{https://doi.org/10.1016/0550-3213(82)90009-8}{\emph{Nucl. Phys. B}
  {\bfseries 195} (1982) 503}.

\bibitem{Haegeman:2011zz}
J.~Haegeman, J.I.~Cirac, T.J.~Osborne, I.~Pizorn, H.~Verschelde and
  F.~Verstraete, \emph{{Time-Dependent Variational Principle for Quantum
  Lattices}}, \href{https://doi.org/10.1103/PhysRevLett.107.070601}{\emph{Phys.
  Rev. Lett.} {\bfseries 107} (2011) 070601}
  [\href{https://arxiv.org/abs/1103.0936}{{\ttfamily 1103.0936}}].

\bibitem{Haegeman:2016gfj}
J.~Haegeman, C.~Lubich, I.~Oseledets, B.~Vandereycken and F.~Verstraete,
  \emph{{Unifying time evolution and optimization with matrix product states}},
  \href{https://doi.org/10.1103/PhysRevB.94.165116}{\emph{Phys. Rev. B}
  {\bfseries 94} (2016) 165116}.

\end{thebibliography}\endgroup

\end{document}